\documentclass[fleqn,twoside]{article}
\usepackage{espcrc2}

\usepackage{graphicx}
\newcommand{\AmS}{{\protect\the\textfont2
  A\kern-.1667em\lower.5ex\hbox{M}\kern-.125emS}}

\title{Scalar meson exchange in $V\rightarrow P^0 P^0\gamma$ decays
       \thanks{UAB--FT--532 report. 
               To appear in the proceedings of the QCD 02
	       High-Energy Physics International Conference
	       in Quantum Chromodynamics,
	       2-9th July 2002 Montpellier (France).}}

\author{R. Escribano\address{Grup de F\'{\i}sica Te\`orica and IFAE, 
        Universitat Aut\`onoma de Barcelona,\\ 
        E-08193 Bellaterra (Barcelona), Spain}%
	\thanks{Work partly supported by the
	        Ministerio de Ciencia y Tecnolog\'{\i}a and FEDER,
		FPA2002-00748,
		and the EU, HPRN-CT-2002-00311, EURIDICE network.}}
       
\begin{document}
    
\begin{abstract}
\noindent
The complementarity between Chiral Perturbation Theory and the
Linear Sigma Model is exploited to study scalar meson exchange in
$V\rightarrow P^0 P^0\gamma$ decays.
The recently reported experimental data on $\phi\rightarrow\pi^0\pi^0\gamma$,
$\phi\rightarrow\pi^0\eta\gamma$ and $\rho\rightarrow\pi^0\pi^0\gamma$
can be satisfactorily accommodated in our framework.
\end{abstract}

\maketitle

\section{INTRODUCTION}
\noindent
The radiative decays of light vector mesons $(V=\rho,\omega,\phi)$
into a pair of neutral pseudoscalars $(P=\pi^0,K^0,\eta)$,
$V\rightarrow P^0P^0\gamma$, are an excellent laboratory for investigating the
nature and extracting the properties of the light scalar meson resonances
$(S=\sigma,a_{0},f_{0})$.
Particularly interesting are the so called \emph{golden processes}, namely
$\phi\rightarrow\pi^0\pi^0\gamma$, $\phi\rightarrow\pi^0\eta\gamma$ and
$\rho\rightarrow\pi^0\pi^0\gamma$, which, as we will see, can provide us with
valuable information on the properties of the $f_{0}(980)$, $a_{0}(980)$ and
$\sigma(500)$, respectively.

\section{EXPERIMENTAL DATA}
\noindent
For $\phi\rightarrow\pi^0\pi^0\gamma$, the first measurements of this decay
have been reported by the SND and CMD-2 Collaborations.
For the branching ratio they obtain
$B(\phi\rightarrow\pi^0\pi^0\gamma)=(1.22\pm 0.12)\times 10^{-4}$
\cite{Achasov:2000ym}
and $(0.92\pm 0.10)\times 10^{-4}$ \cite{Akhmetshin:1999di},
for $m_{\pi\pi}>700$ MeV in the latter case.
More recently, the KLOE Collaboration has measured
$B(\phi\rightarrow\pi^0\pi^0\gamma)=(1.09\pm 0.06)\times 10^{-4}$
\cite{Aloisio:2002bt}.
In all the cases, the spectrum is clearly peaked at $m_{\pi\pi}\simeq 970$ MeV,
as expected from an important $f_{0}(980)$ contribution.

\noindent
For $\phi\rightarrow\pi^0\eta\gamma$, the branching ratios measured by the
SND, CMD-2 and KLOE Collaborations are
$B(\phi\rightarrow\pi^0\eta\gamma)=(8.8\pm 1.7)\times 10^{-5}$
\cite{Achasov:2000ku},
$(9.0\pm 2.6)\times 10^{-5}$ \cite{Akhmetshin:1999di}, and
$B(\phi\rightarrow\pi^0\eta\gamma)=(8.51\pm 0.76)\times 10^{-5}$
$(\eta\rightarrow\gamma\gamma)$ and
$(7.96\pm 0.72)\times 10^{-5}$ $(\eta\rightarrow\pi^+\pi^-\pi^0)$
\cite{Aloisio:2002bs}.
Again, in all the cases, the observed mass spectrum shows a
significant enhancement at large $\pi^0\eta$ invariant mass that is
interpreted as a manifestation of the dominant contribution of the
$a_{0}(980)\gamma$ intermediate state.

\noindent
For $\rho\rightarrow\pi^0\pi^0\gamma$, the SND Coll.~has measured
$B(\rho\rightarrow\pi^0\pi^0\gamma)=(4.1^{+1.0}_{-0.9}\pm 0.3)\times 10^{-5}$
\cite{Achasov:2002jv}.
This value can be explained by means of a significant contribution of the
$\sigma(500)\gamma$ intermediate state together with the well-known
$\omega\pi$ contribution.

\section{THEORETICAL FRAMEWORK}
\noindent
A first attempt to explain the $V\rightarrow P^0P^0\gamma$ decays was done in
Ref.~\cite{Bramon:1992kr} using the vector meson dominance (VMD) model.
In this framework, the $V\rightarrow P^0P^0\gamma$ decays proceed through the
decay chain $V\rightarrow VP^0\rightarrow P^0P^0\gamma$.
The calculated branching ratios
$B_{\phi\rightarrow\pi^0\pi^0\gamma}^{\rm VMD}=1.2\times 10^{-5}$,
$B_{\phi\rightarrow\pi^0\eta\gamma}^{\rm VMD}=5.4\times 10^{-6}$ and
$B_{\rho\rightarrow\pi^0\pi^0\gamma}^{\rm VMD}=1.1\times 10^{-5}$
\cite{Bramon:1992kr}
are found to be substantially smaller than the experimental results.
Later on, the $V\rightarrow P^0P^0\gamma$ decays were studied in a 
Chiral Perturbation Theory (ChPT) context enlarged to included on-shell vector
mesons \cite{Bramon:1992ki}. In this formalism,
$B_{\phi\rightarrow\pi^0\pi^0\gamma}^\chi=5.1\times 10^{-5}$,
$B_{\phi\rightarrow\pi^0\eta\gamma}^\chi=3.0\times 10^{-5}$ and
$B_{\rho\rightarrow\pi^0\pi^0\gamma}^\chi=9.5\times 10^{-6}$.
Taking into account both chiral and VMD contributions, one finally obtains
$B_{\phi\rightarrow\pi^0\pi^0\gamma}^{{\rm VMD}+\chi}=6.1\times 10^{-5}$,
$B_{\phi\rightarrow\pi^0\eta\gamma}^{{\rm VMD}+\chi}=3.6\times 10^{-5}$ and
$B_{\rho\rightarrow\pi^0\pi^0\gamma}^{{\rm VMD}+\chi}=2.6\times 10^{-5}$
\cite{Bramon:1992ki}, which are still below the experimental results.
Additional contributions are thus certainly required and the most natural 
candidates are the contributions coming from the exchange of scalar resonances.
A first model including the scalar resonances explicitly is the
\emph{no structure model}, where the $V\rightarrow P^0P^0\gamma$ decays
proceed through the decay chain $V\rightarrow S\gamma\rightarrow P^0P^0\gamma$
and the coupling $VS\gamma$ is considered as pointlike.
This model is ruled out by experimental data on
$\phi\rightarrow\pi^0\pi^0\gamma$ decays \cite{Achasov:2000ym}.
A second model is the \emph{kaon loop model} \cite{Achasov:1987ts},
where the initial vector decays into a pair of charged kaons that,
after the emission of a photon, rescatter into a pair of neutral pseudoscalars
through the exchange of scalar resonances.

\noindent
The previous two models include the scalar resonances {\it ad hoc},
and the pseudoscalar rescattering amplitudes are not chiral invariant.
This problem is solved in the next two models which are based not only on the
\emph{kaon loop model} but also on chiral symmetry.
The first one is the Unitarized Chiral Perturbation Theory (UChPT) where the 
scalar resonances are generated dynamically by unitarizing the one-loop
pseudoscalar amplitudes.
In this approach,
$B_{\phi\rightarrow\pi^0\pi^0\gamma}^{\rm U\chi PT}=8\times 10^{-5}$,
$B_{\phi\rightarrow\pi^0\eta\gamma}^{\rm U\chi PT}=8.7\times 10^{-5}$ and
$B_{\rho\rightarrow\pi^0\pi^0\gamma}^{\rm U\chi PT}=1.4\times 10^{-5}$
\cite{Marco:1999df}.
The second model is the Linear Sigma Model (L$\sigma$M), a well-defined
$U(3)\times U(3)$ chiral model which incorporates {\it ab initio} the
pseudoscalar and scalar mesons nonets.
The advantage of the L$\sigma$M is to incorporate explicitly the effects of
scalar meson poles while keeping the correct behaviour at low invariant masses
expected from ChPT.

\noindent
In the next three sections, we discuss the scalar contributions to the
$\phi\rightarrow\pi^0\pi^0\gamma$, $\phi\rightarrow\pi^0\eta\gamma$ and
$\rho\rightarrow\pi^0\pi^0\gamma$  decays in the framework of the L$\sigma$M.

\section{$\phi\rightarrow\pi^0\pi^0\gamma$}
\noindent
The scalar contribution to this process is driven by the decay chain
$\phi\rightarrow K^+K^-(\gamma)\rightarrow\pi^0\pi^0\gamma$.
The contribution from pion loops is known to be negligible due to the Zweig rule.
The amplitude for 
$\phi(q^\ast,\epsilon^\ast)\rightarrow\pi^0(p)\pi^0(p^\prime)\gamma(q,\epsilon)$
is given by \cite{Bramon:2002iw}
\begin{equation}
\label{Aphipi0pi0gamma}
{\cal A}=\frac{eg_{s}}{2\pi^2 m^2_{K^+}}\,\{a\}\,L(m^2_{\pi^0\pi^0})
\times{\cal A}_{K^+ K^-\rightarrow\pi^0\pi^0}^{\mbox{\scriptsize L$\sigma$M}}\, ,
\end{equation} 
where
$\{a\}=(\epsilon^\ast\cdot\epsilon)\,(q^\ast\cdot q)-
       (\epsilon^\ast\cdot q)\,(\epsilon\cdot q^\ast)$, 
$m^2_{\pi^0\pi^0}\equiv s$ is the dipion invariant mass and $L(m^2_{\pi^0\pi^0})$
is a loop integral function.
The $\phi K\bar K$ coupling constant $g_{s}$ takes the value $|g_s|\simeq 4.6$
to agree with $\Gamma_{\phi\rightarrow K^+K^-}^{\rm exp}= 2.19$ MeV
\cite{Groom:2000in}.
\begin{figure}[t]
\centerline{\includegraphics[width=0.45\textwidth]{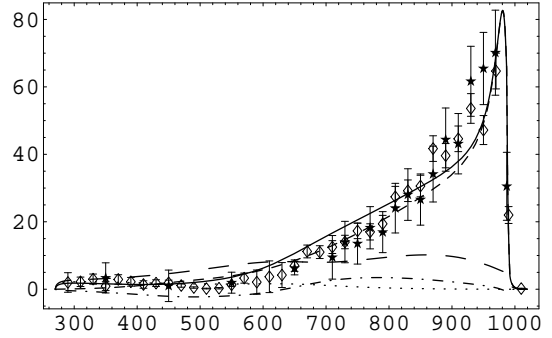}} 
\vspace{-0.7cm}
\caption{\small
$dB(\phi\rightarrow\pi^0\pi^0\gamma)/dm_{\pi^0\pi^0} \times 10^8$
(in MeV$^{-1}$) {\it versus} $m_{\pi^0\pi^0}$ (in MeV). 
The dashed, dotted and dot-dashed lines correspond to the contributions from
the L$\sigma$M, VMD and their interference, respectively. 
The solid line is the total result.
The long-dashed line is the chiral loop prediction.
Experimental data are taken from Ref.~\protect\cite{Achasov:2000ym} (solid star)
and Ref.~\protect\cite{Aloisio:2002bt} (open diamond).}
\label{dBdmallcontribF}
\vspace{-0.7cm}
\end{figure}

\noindent
The $K^+ K^-\rightarrow\pi^0\pi^0$ amplitude in the L$\sigma$M is
\begin{equation}
\label{AKKpipiLsM}
\begin{array}{l}
{\cal A}_{K^+K^-\rightarrow\pi^0\pi^0}^{\mbox{\scriptsize L$\sigma$M}}=
\frac{m^2_{\pi}-s/2}{2f_\pi f_K}\\[1ex]
\quad+\frac{s-m^2_{\pi}}{2f_\pi f_K}
\times\left[
\frac{m^2_K-m^2_{\sigma}}{D_{\sigma}(s)}
{\rm c}\phi_S({\rm c}\phi_S-\sqrt{2}\,{\rm s}\phi_S)\right.\\
\qquad\qquad\quad\left.+
\frac{m^2_K-m^2_{f_0}}{D_{f_0}(s)}
{\rm s}\phi_S({\rm s}\phi_S+\sqrt{2}\,{\rm c}\phi_S)
\right]\, ,
\end{array}
\end{equation}
where $D_{S}(s)$ are the $S=\sigma, f_0$ propagators,
$\phi_S$ is the scalar mixing angle in the flavour basis
and $({\rm c}\phi_S, {\rm s}\phi_S)\equiv (\cos\phi_S, \sin\phi_S)$.
A Breit-Wigner propagator is used for the $\sigma$, while for the $f_{0}$
a complete one-loop propagator is preferable
\cite{Achasov:1987ts,Escribano:2002iv}.
For $m_{S}\rightarrow\infty\ (S=\sigma,f_{0})$,
the L$\sigma$M amplitude (\ref{AKKpipiLsM}) reduces to the corresponding
ChPT amplitude, $s/(4f_{\pi}f_{K})$,
and is thus expected to account for the lowest part of the $\pi\pi$ spectrum.
In addition, the presence of the scalar propagators in Eq.~(\ref{AKKpipiLsM})
should be able to reproduce the effects of the $f_{0}$ (and the $\sigma$)
pole(s) at higher $\pi\pi$ invariant mass values.
This complementarity between ChPT and the L$\sigma$M makes the whole analysis
quite reliable.

\noindent
The final results for ${\cal A}(\phi\rightarrow\pi^0\pi^0\gamma)$ are then the
sum of the L$\sigma$M contribution in Eq.~(\ref{Aphipi0pi0gamma}) plus the VMD
contribution that can be found in Ref.~\cite{Bramon:2002iw}.
The $\pi^0\pi^0$ invariant mass distribution,
with the separate contributions from the L$\sigma$M, VMD and their interference,
as well as the total result, are shown in Fig.~\ref{dBdmallcontribF}.
We use $m_\sigma=478$ MeV and $\Gamma_\sigma=324$ MeV from 
Ref.~\cite{Aitala:2001xu}, $m_{f_{0}}=985$ MeV and $\phi_{S}=-9^\circ$
\cite{Bramon:2002iw}.
Notice that the contribution of the $\sigma$ to this process is suppressed
since $g_{\sigma KK}\propto (m^2_{\sigma}-m^2_{K})\simeq 0$ for
$m_{\sigma}\simeq m_{K}$
---by contrast, the chiral loop prediction shows no suppression in the region
$m_{\pi\pi}\simeq 500$ MeV, see Fig.~\ref{dBdmallcontribF}.

\noindent
Integrating the $\pi^0\pi^0$ invariant mass distribution over the whole physical
region one finally obtains
$B(\phi\rightarrow\pi^0\pi^0\gamma)=1.16\times 10^{-4}$.
The shape of the $\pi\pi$ mass spectrum and the branching ratio are in
agreement with the experimental results.
However, both predictions are very sensitive to the values of the $f_{0}$ mass
and the scalar mixing angle
(this latter because of $g_{f_{0}\pi\pi}\propto\sin\phi_{S}$).
Consequently, the $\phi\rightarrow\pi^0\pi^0\gamma$ decay could be used to 
extract valuable information on these parameters.

\section{$\phi\rightarrow\pi^0\eta\gamma$}
\begin{figure}[t]
\centerline{\includegraphics[width=0.45\textwidth]{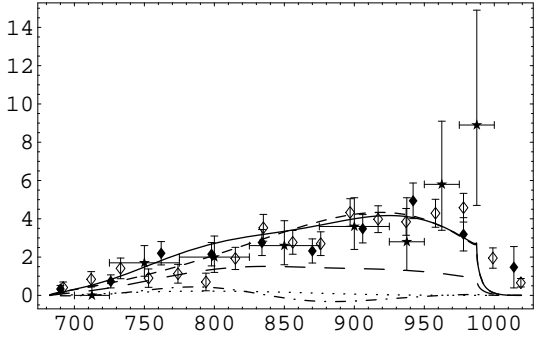}} 
\vspace{-0.7cm}
\caption{\small
$dB(\phi\rightarrow\pi^0\eta\gamma)/dm_{\pi^0\eta} \times 10^7$
(in MeV$^{-1}$) {\it versus} $m_{\pi^0\eta}$ (in MeV). 
The curves follow the same conventions as in Fig.~\protect\ref{dBdmallcontribF}.
Experimental data are taken from Ref.~\protect\cite{Achasov:2000ku} (solid star)
and Ref.~\protect\cite{Aloisio:2002bs}:
(open diamond)  from $\eta\rightarrow\gamma\gamma$ and
(solid diamond) from $\eta\rightarrow\pi^+\pi^-\pi^0$.}
\label{dBdmpietaallcontribF}
\vspace{-0.7cm}
\end{figure}

\noindent
The scalar contribution to the $\phi\rightarrow\pi^0\eta\gamma$ decay
is identical to that of $\phi\rightarrow\pi^0\pi^0\gamma$ with the replacement
of ${\cal A}_{K^+K^-\rightarrow\pi^0\pi^0}^{\mbox{\scriptsize L$\sigma$M}}$
by ${\cal A}_{K^+K^-\rightarrow\pi^0\eta}^{\mbox{\scriptsize L$\sigma$M}}$
in Eq.~(\ref{AKKpipiLsM}).
This latter amplitude is written as
\begin{equation}
\label{AKKpietaLsM}
\begin{array}{l}
{\cal A}_{K^+K^-\rightarrow\pi^0\eta}^{\mbox{\scriptsize L$\sigma$M}}
=\frac{m^2_{\eta}+m^2_{\pi}-s}{4f_\pi f_K}
({\rm c}\phi_{P}-\sqrt{2}\,{\rm s}\phi_{P})\\[1ex]
\qquad\qquad\qquad
+\frac{s-m^2_{\eta}}{2f_\pi f_K}
\frac{m^2_K-m^2_{a_0}}{D_{a_0}(s)}\,{\rm c}\phi_{P}\, ,
\end{array}
\end{equation}
where $\phi_{P}$ is the pseudoscalar mixing angle and
$D_{a_0}(s)$ the complete one-loop $a_{0}$ propagator \cite{Achasov:1987ts}.

\noindent
The separate contributions to the $\pi^0\eta$ invariant mass distribution,
and the total result, are shown in Fig.~\ref{dBdmpietaallcontribF}.
The chiral loop prediction is also included for comparison.
We use $m_{a_{0}}=984.8$ MeV \cite{Groom:2000in} and
$\phi_{P}=41.8^\circ$ \cite{Aloisio:2002vm}.
Integrating the $\pi^0\eta$ invariant mass spectrum one obtains
$B(\phi\rightarrow\pi^0\eta\gamma)=8.3\times 10^{-5}$.
The $\pi^0\eta$ mass spectrum and the branching ratio are in fair agreement
with experimental results and with previous phenomenological estimates
\cite{Bramon:2000vu,Escribano:2000fs}.

\section{$\rho\rightarrow\pi^0\pi^0\gamma$}
\begin{figure}[t]
\centerline{\includegraphics[width=0.45\textwidth]{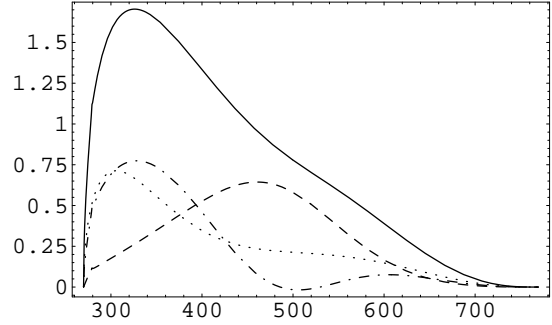}} 
\vspace{-0.7cm}
\caption{\small
$dB(\rho\rightarrow\pi^0\pi^0\gamma)/dm_{\pi^0\pi^0}\times 10^7$ (in MeV$^{-1}$)
{\it versus} $m_{\pi^0\pi^0}$ (in MeV). 
The curves follow the same conventions as in Fig.~\protect\ref{dBdmallcontribF}.
The reference values $m_\sigma=478$ MeV and $\Gamma_\sigma=324$ MeV
\protect\cite{Aitala:2001xu} have been used.}
\label{dBdmrhopipiallcontribF}
\vspace{-0.7cm}
\end{figure}

\noindent
The scalar contribution to this process is mainly driven by the decay mechanism
$\rho\rightarrow\pi^+\pi^-(\gamma)\rightarrow\pi^0\pi^0\gamma$
\cite{Bramon:1992ki},
and the amplitude is given by \cite{Bramon:2001un}
\begin{equation}
\label{Arhopi0pi0gamma}
{\cal A}=\frac{-eg}{\sqrt{2}\pi^2 m^2_{\pi^+}}\,\{a\}\,L(m^2_{\pi^0\pi^0})\times
{\cal A}_{\pi^+\pi^-\rightarrow\pi^0\pi^0}^{\mbox{\scriptsize L$\sigma$M}}\, ,
\end{equation} 
where the $\rho\pi\pi$ coupling constant $g$ is fixed to $|g|=4.27$ to agree with
$\Gamma_{\rho\rightarrow\pi^+\pi^-}^{\rm exp}=150.2$ MeV
---in the good $SU(3)$ limit one should have $|g|=|g_{s}|$.

\noindent
The $\pi^+\pi^-\rightarrow\pi^0\pi^0$ amplitude in the L$\sigma$M is
\begin{equation} 
\label{ApipipipiLsM}
\begin{array}{l}
{\cal A}_{\pi^+\pi^-\rightarrow\pi^0\pi^0}^{\mbox{\scriptsize L$\sigma$M}}=   
\frac{s-m^2_\pi}{f_\pi^2}\\[1ex]
\qquad\quad\times  
\left(\frac{m^2_\pi-m^2_\sigma}{D_\sigma(s)}\,{\rm c}^2\phi_S+ 
      \frac{m^2_\pi-m^2_{f_0}}{D_{f_0}(s)}\,{\rm s}^2\phi_S\right)\, , 
\end{array}
\end{equation}
which reduces to the ChPT amplitude, $(s-m^2_\pi)/f_\pi^2$,
in the limit $m_{\sigma,f_{0}}\rightarrow\infty$.

\noindent
The different contributions to the $\pi^0\pi^0$ invariant mass distribution
are shown in Fig.~\ref{dBdmrhopipiallcontribF}.
Integrating the $\pi^0\pi^0$ mass spectrum one obtains
$B(\rho\rightarrow\pi^0\pi^0\gamma)=3.8\times 10^{-5}$,
in agreement with the experimental result.

\begin{figure}[t] 
\centerline{\includegraphics[width=0.45\textwidth]{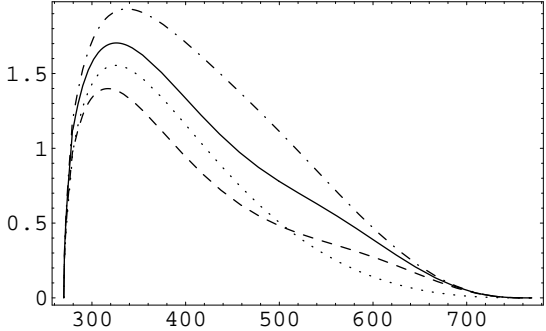}}  
\vspace{-0.7cm}
\caption{\small 
$dB(\rho\rightarrow\pi^0\pi^0\gamma)/dm_{\pi^0\pi^0}\times 10^7$ (in MeV$^{-1}$)
{\it versus} $m_{\pi^0\pi^0}$ (in MeV). 
The various predictions are for:  
$m_\sigma = 478$ MeV and $\Gamma_\sigma = 324$ MeV \protect\cite{Aitala:2001xu}
(solid line);   
$m_\sigma = 478$ MeV \protect\cite{Aitala:2001xu} and
$\Gamma_\sigma^{\mbox{\scriptsize L$\sigma$M}} = 256$ MeV (dot-dashed line); and  
$m_\sigma = 555$ MeV and $\Gamma_\sigma = 540$ MeV \protect\cite{Asner:2000kj}
(dashed line).  
The chiral loop prediction is also included for comparison (dotted line).} 
\label{figrhoall} 
\vspace{-0.8cm}
\end{figure}

\noindent
The scalar contribution to $\rho\rightarrow\pi^0\pi^0\gamma$ is mainly driven
by the $\sigma(500)$ for kinematical reasons.
In order to show the sensitivity of our treatment on the parameters of the  
$\sigma$ we have plotted in Fig.~\ref{figrhoall} our final predictions
for various values of $m_\sigma$ and $\Gamma_\sigma$. 
Taking now the values $m_\sigma=478$ MeV \cite{Aitala:2001xu} 
and $\Gamma_{\sigma}=256$ MeV, as required by the L$\sigma$M, one finds   
$B(\rho\rightarrow\pi^0\pi^0\gamma)=4.7\times 10^{-5}$. 
The prediction for $m_\sigma=555$ MeV and $\Gamma_{\sigma}=540$ MeV
\cite{Asner:2000kj} is $2.8\times 10^{-5}$, well below the SND result.
The smallness of the former value disfavours a broad ${\sigma}$
while the smallness of the chiral loop contribution $2.9\times 10^{-5}$
confirms the need of the effects of a moderately narrow ${\sigma}$.  

\section{CONCLUSIONS}
\noindent
{\it i)}
The \emph{golden processes}, $\phi\rightarrow\pi^0\pi^0\gamma$,
$\phi\rightarrow\pi^0\eta\gamma$ and $\rho\rightarrow\pi^0\pi^0\gamma$
have been shown to be very useful to extract relevant information on the
properties of the $f_{0}(980)$, $a_{0}(980)$ and $\sigma(500)$, respectively.

\noindent
{\it ii)}
The complementary between ChPT and the L$\sigma$M is used to parametrize the
needed scalar amplitudes. This guarantees the appropriate behaviour at low
dimeson invariant masses but also allows to include the effects of the 
scalar meson poles.

\noindent
{\it iii)}
The L$\sigma$M predictions for the invariant mass spectra and their
respective branching ratios are compatible with experimental data.

\noindent
{\it iv)}
The prediction for $\phi\rightarrow\pi^0\pi^0\gamma$ is dominated by
$f_{0}(980)$ exchange and is strongly dependent on the values of $m_{f_{0}}$
and $\phi_{S}$.
For the preferred values $m_{f_{0}}=985$ MeV and $\phi_{S}=-9^\circ$,
one obtains $B(\phi\rightarrow\pi^0\pi^0\gamma)=1.16\times 10^{-4}$.

\noindent
{\it v)}
$\phi\rightarrow\pi^0\eta\gamma$ is dominated by $a_{0}(980)$ exchange.
For the values $m_{a_{0}}=984.8$ MeV and $\phi_{P}=41.8^\circ$,
one obtains $B(\phi\rightarrow\pi^0\eta\gamma)=8.3\times 10^{-5}$.

\noindent
{\it vi)}
Experimental data on $\rho\rightarrow\pi^0\pi^0\gamma$ decays
seem to prefer a low mass and moderately narrow $\sigma(500)$.
For the reference values $m_{\sigma}=478$ MeV and $\Gamma_{\sigma}=324$ MeV,
one obtains $B(\rho\rightarrow\pi^0\pi^0\gamma)=3.8\times 10^{-5}$.

\end{document}